\documentclass[mathleft]{an}
\usepackage{graphicx}
\usepackage{times}
\begin{document}

\Pagespan{0}{}
\Yearpublication{2014}%
\Yearsubmission{2014}%
\Month{}%
\Volume{}%
\Issue{}%

\title{The orbital evolution of a passive high-orbit fragment with large surface area.}

\author{
A.A. Bazyey 
\inst{1}
\fnmsep\thanks{Corresponding author:
\email{$o.bazyey@onu.edu.ua$}
\newline
}
\and
N.V. Bazyey 
\inst{1}
\and 
V.I. Kashuba 
\inst{1}
\and 
S.G. Kashuba 
\inst{1}
\and 
V.V. Kouprianov 
\inst{2}
\and 
I.E. Molotov 
\inst{3}
\and 
Z.N. Khutorovsky 
\inst{4}
\and 
L.G. Tsybizova 
\inst{1}
}
\titlerunning{The orbital evolution of a passive high-orbit fragment}
\authorrunning{Bazyey et al.}
\institute{Astronomical Observatory of I.I. Mechnikov 
Odessa National University, Marazlievska St. 1v, Odessa 65014, Ukraine
\and 
The Main Astronomical Observatory of the Russian Academy of Sciences, Pulkovo, 
Pulkovo highway 65, Saint-Petersburg 196140, Russia 
\and 
Keldysh Institute of Applied Mathematics of the Russian Academy of Sciences, 
Miusska square 4, Moscow 125047, Russia
\and
JSC Interstate joint-stock corporation Vympel, the 4th Vosmoye Marta St. 3, Moscow 125167, 
Russia
}
\received{}
\accepted{}
\publonline{later}
\abstract{The observation data for artificial celestial body 43096, which had been obtained 
during 2006-2012 within the framework of international project "The Scientific Network of 
Optical Instruments for Astrometric and Photometric Observations" - International Scientific Optical Network 
(ISON), were processed. The Keplerian elements and state vector as of 24 November 2006 01:55:50.76 UTC 
were determined.\\
The numerical integration of the motion equations was performed accounting for the perturbations due 
to the polar flattening of the Earth, Moon and Sun, as well as the solar radiation pressure.\\
Based on the numerical model of a motion in the near-Earth space that accounts for only the most 
powerful perturbations, a new method for de-orbiting artificial celestial bodies from high altitudes 
is suggested.\\
For the first time such a considerable amount of data over long time intervals was gathered for the 
objects with high area-to-mass ratio that enabled to determine their specific characteristics.}
\maketitle

\section{Introduction}
Today there are tens of thousands of artificial celestial bodies in the near-Earth space. Most of them 
belong to the space debris as such worn-out artificial satellites or their fragments. Such celestial 
bodies can remain in high orbits essentially indefinitely. Their motion is subjected to the perturbations 
by the Moon and Sun, as well as by the asymmetry of the Earth's gravitational field. The high-orbit objects 
are monitored using optical telescopes. The international project "The Scientific Network of Optical 
Instruments for Astrometric and Photometric Observations" - International Scientific Optical Network 
(ISON, Molotov et al. 2009)  contributes the most.\\
This paper describes the processing of the positional observation data of one of passive high-orbit celestial 
bodies. Based on the obtained results, a new method for de-orbiting of worn-out artificial satellites from 
the geostationary orbits in the near-Earth space to lower altitudes is proposed. 

\section{Observations}
We selected fragment 43096, which was detected with the ESA Space Debris 1-m Telescope located on the island 
of Tenerife, Spain, by Thomas Schildknecht's team during their cooperation with the ISON project (Volvach et al. 2006). 
The fragment's reference corresponds to the number in the Keldysh Institute of Applied Mathematics of the Russian 
Academy of Sciences database. This fragment is interesting by its high area-to-mass ratio (HAMR). When describing 
changes in its orbiting, it is necessary to account for significant perturbations due to the radiation pressure 
in addition to those by gravity. Perturbations due to the solar radiation pressure tend to be periodic.\\
We processed observation data for the indicated fragment, which had been obtained by the ISON network during 
2006-2012 within the framework of the Pulkovo Cooperation of Optical Observers (PulCOO) programme in the following 
observation stations: Tenerife (Zeiss-1000), Zimmerwald (ZIMLAT), Crimean Astronomical Observatory (AT-64 and 
RST-220), Nauchniy village (Zeiss-600), Mondy village (infrared reflector ART-33), Mount Maydanak (Zeiss-600), 
Mayaki village (the Ritchey-Chretien telescope RC-600), Gissar (reflector ART-8), Terskol Peak (Zeiss-2000), 
Abastumani (AC-32), Andrushevka (S-600), Ussuriysk (ORI-50), Artem (ORI-25).

\section{Results of calculations} 
A total of 226 series of observations conducted from 18 November 2006 to 16 June 2012 were processed. Each series 
averaged to 20-30 measurements of the topocentric right ascensions, declinations and UTC time references. 
The Keplerian elements were determined for each series by Laplace's method with the subsequent refining by 
the 6-parameter iteration method. The accuracy of the orbital elements was estimated using the residual errors 
representing differences of the observed positions of the fragment from the predicted ones. The computation 
procedure is specified in Bazyey et al. (2005) and Escobal (1970). The least errors in orbital element determination 
were obtained for the following series of observations:
\\
24 November 2006 01:09:56.87 (Tenerife)
\\
$p = (6.4666 \pm 0.0002)$ equatorial radius
\\
$e = (0.06681 \pm 0.00002)$
\\
$\omega = (261.16 \pm 0.02)^\circ$
\\
$\Omega =  (321.588 \pm 0.002)^\circ$
\\
$i = (9.0212 \pm 0.0003)^\circ$
\\
$M_{0} = (242.48 \pm 0.03) ^\circ$
\\
\\
08 February 2008 23:35:46.77 (Tenerife)
\\
$p = (6.4816 \pm 0.0002)$ equatorial radius
\\
$e = (0.04697 \pm 0.00001)$
\\
$\omega = (295.38 \pm 0.02)^\circ$
\\
$\Omega = (319.765 \pm 0.001)^\circ$
\\
$i = (8.2431 \pm 0.0002)^\circ$
\\
$M_{0} = (240.16 \pm 0.02)^\circ$
\\

\begin{figure}
\resizebox{70 mm}{45 mm}
{\includegraphics {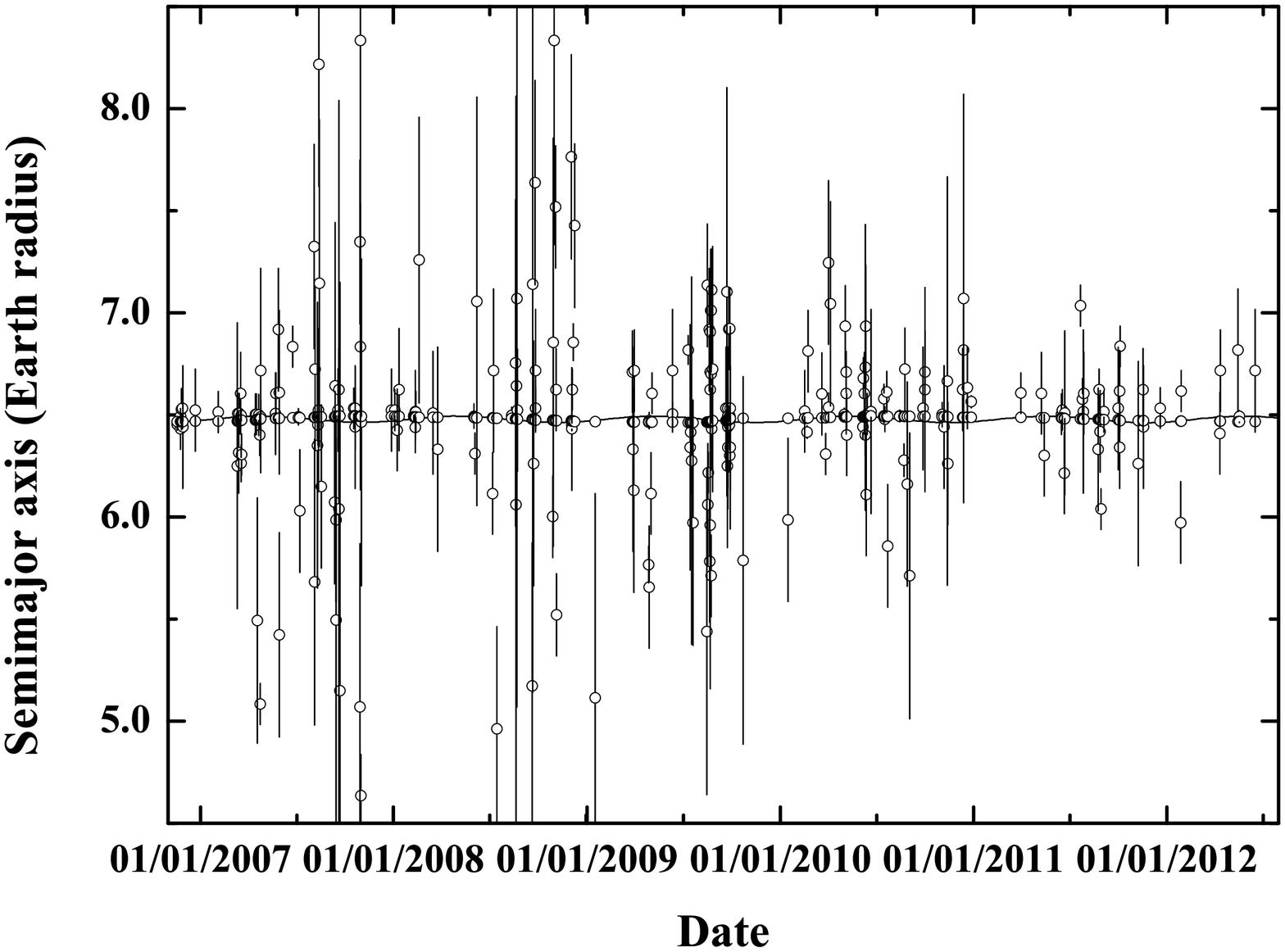}}
\resizebox{70 mm}{45 mm}
{\includegraphics {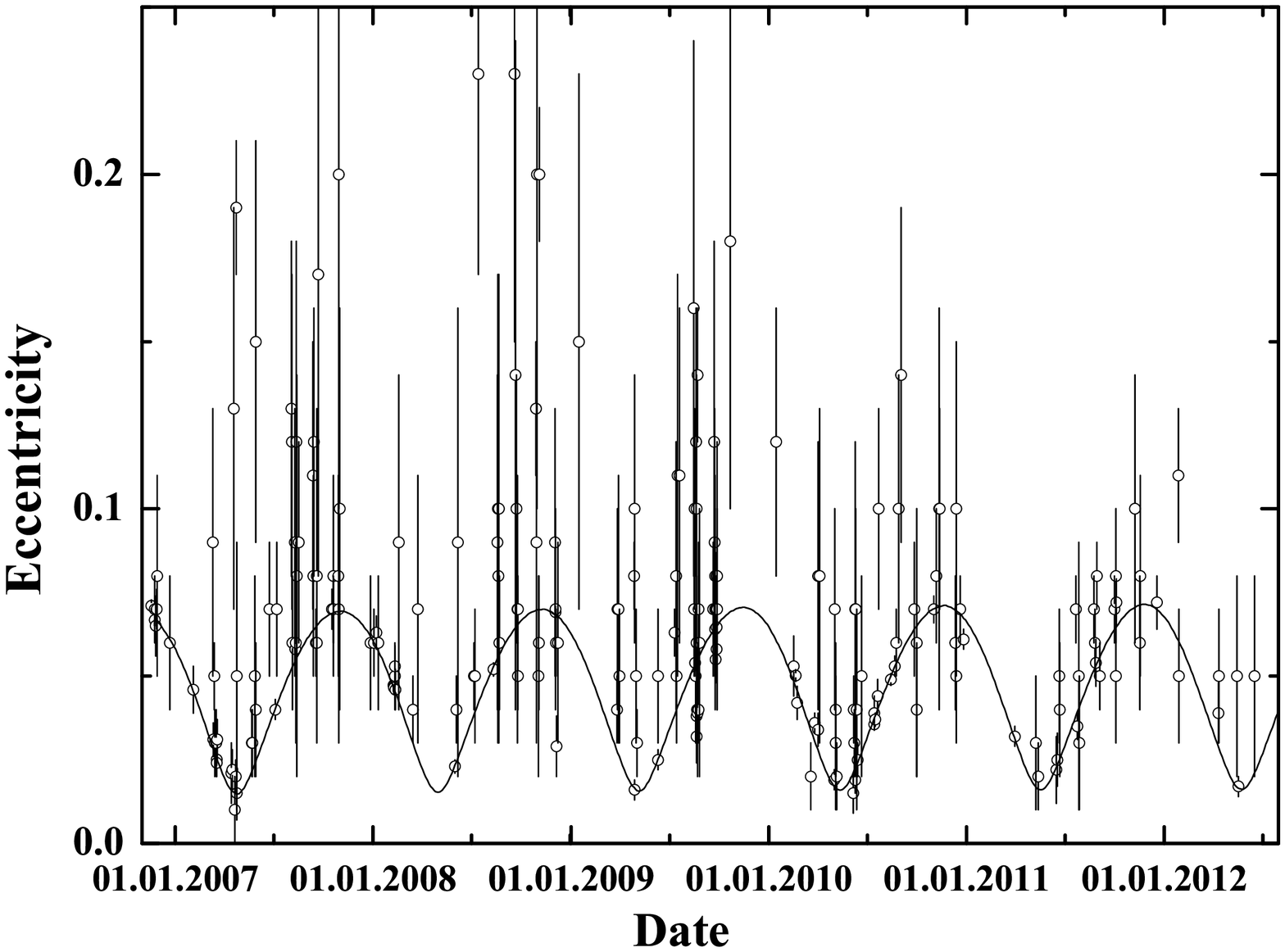}}
\resizebox{70 mm}{45 mm}
{\includegraphics {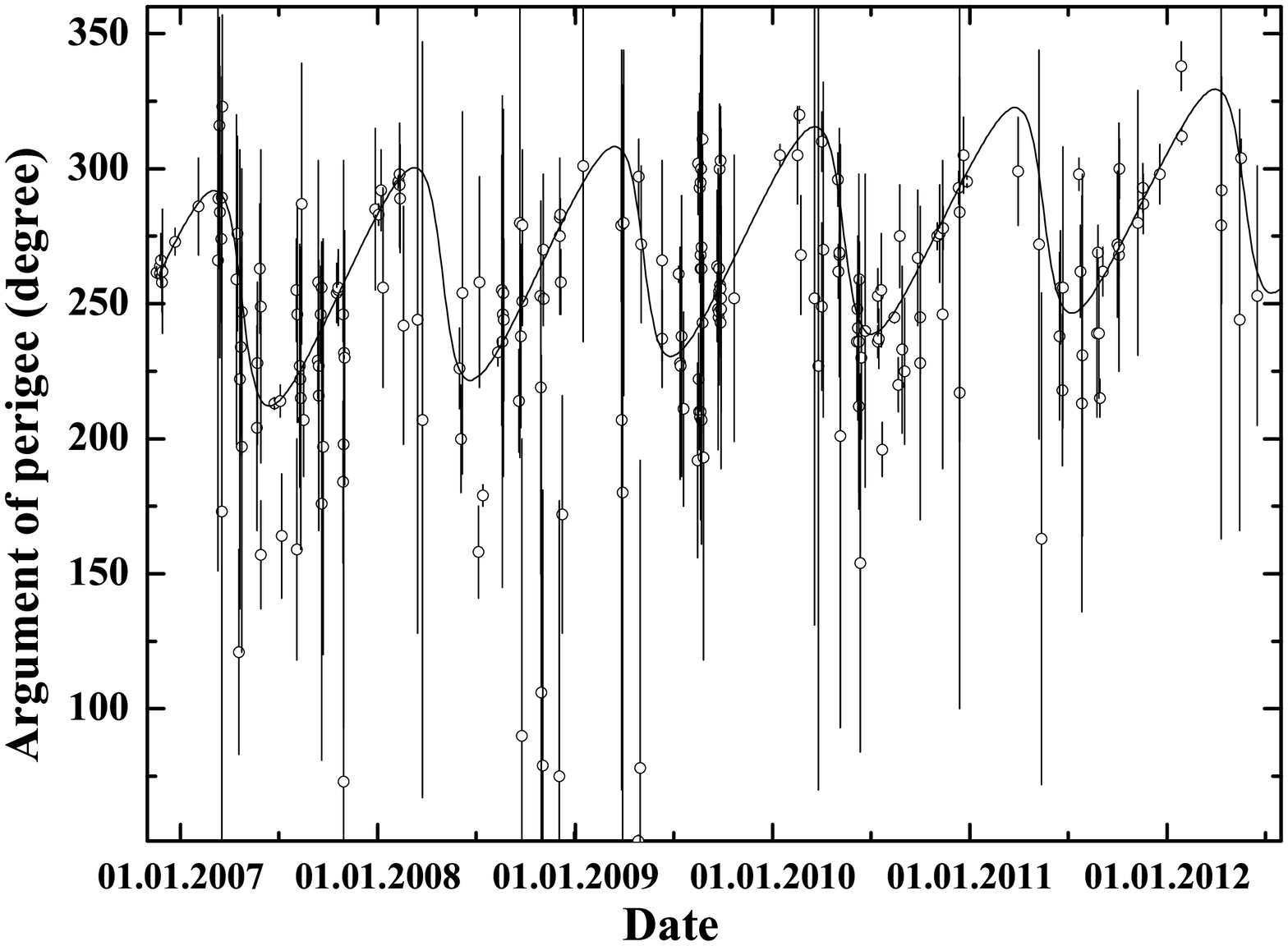}}
\resizebox{70 mm}{45 mm}
{\includegraphics {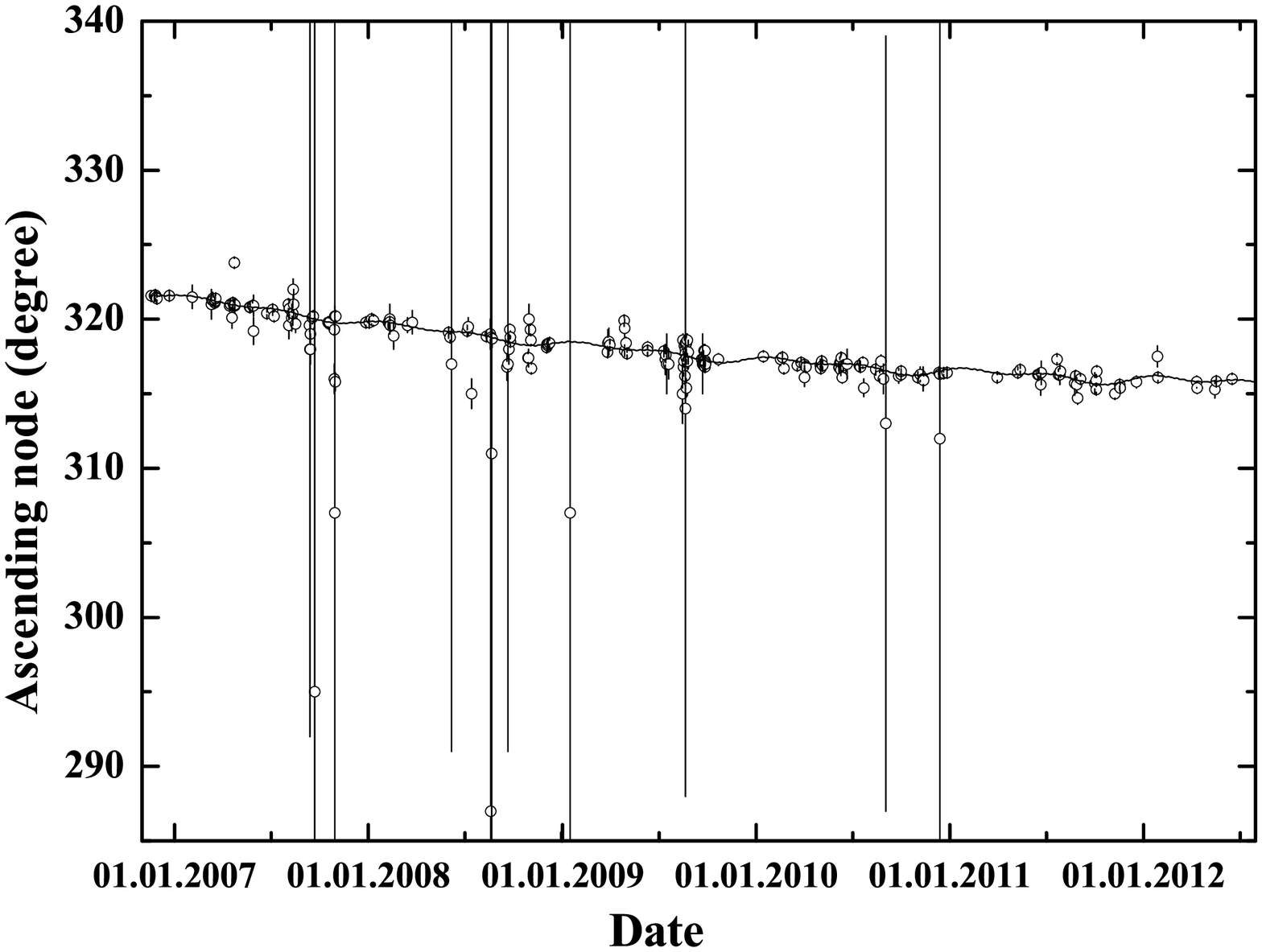}}
\resizebox{70 mm}{45 mm}
{\includegraphics {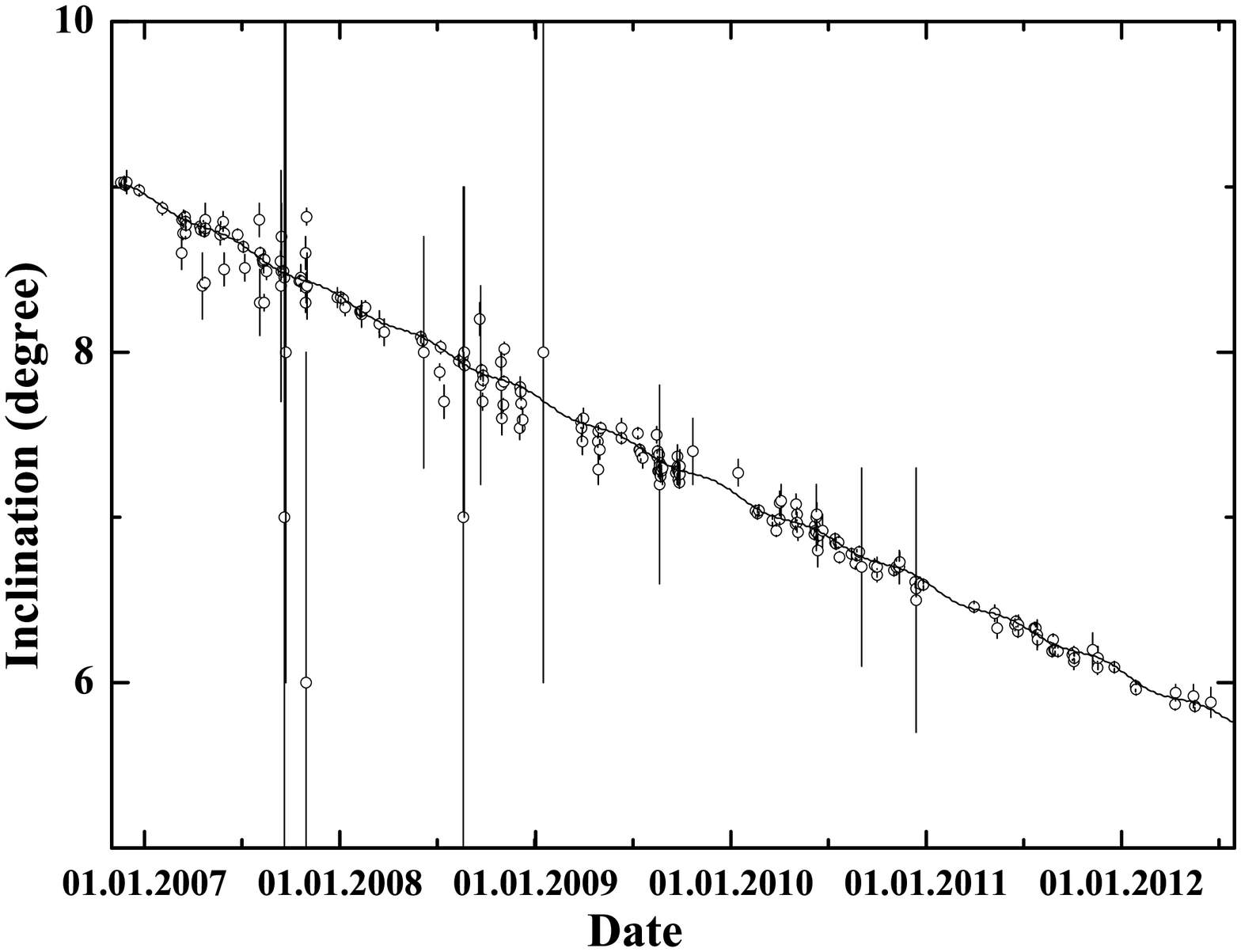}}
\caption{The orbital elements of fragment 43096 in 2006-2012. The observed values are marked as dots, and the 
solid line represents the computational simulation solution.}
\label{fig1}
\end{figure}

\begin{figure}
\resizebox{70 mm}{45 mm}
{\includegraphics {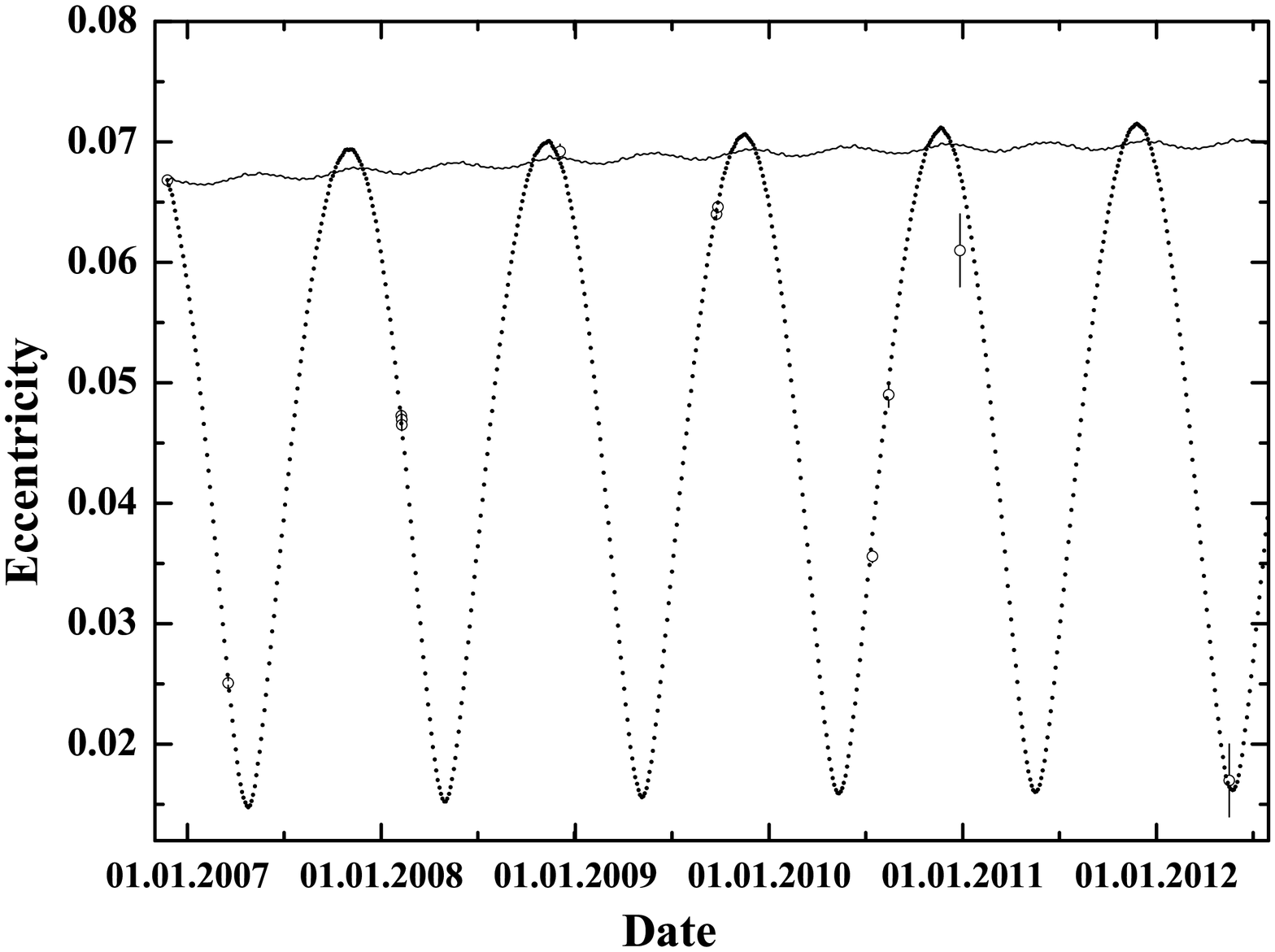}}
\resizebox{70 mm}{45 mm}
{\includegraphics {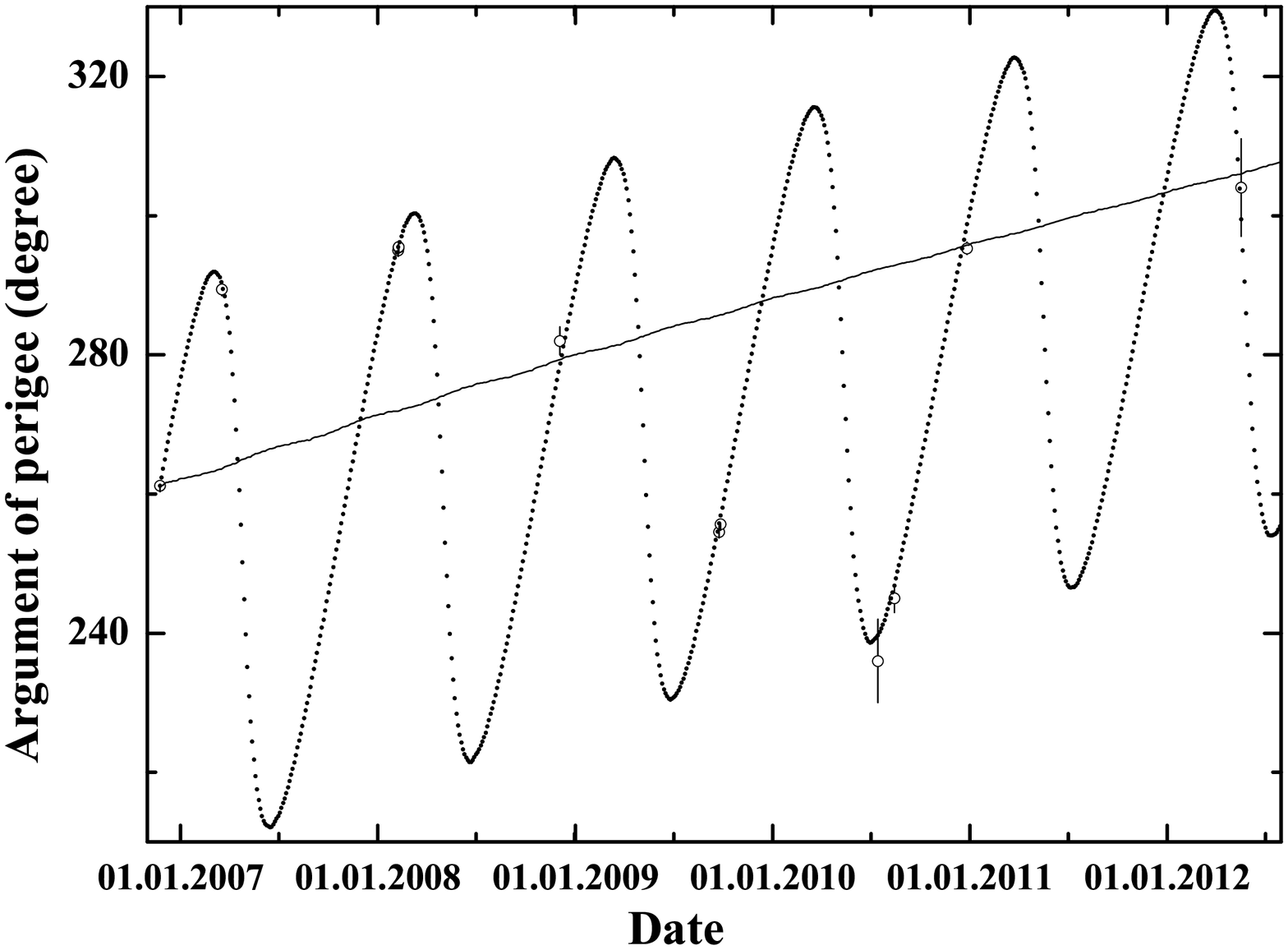}}
\caption{A selection from the orbital elements of fragment 43096. The values with the least absolute error in the 
eccentricity determination (less than 0.002) are marked as dots. The orbital element changes, which were obtained 
by the numerical integration of the motion equations, are represented as the solid line for the case of not allowing 
for the solar radiation pressure and as the dashed line when the radiation pressure is taken into account.}
\label{fig2}
\end{figure}

\begin{figure}
\resizebox{70 mm}{45 mm}
{\includegraphics {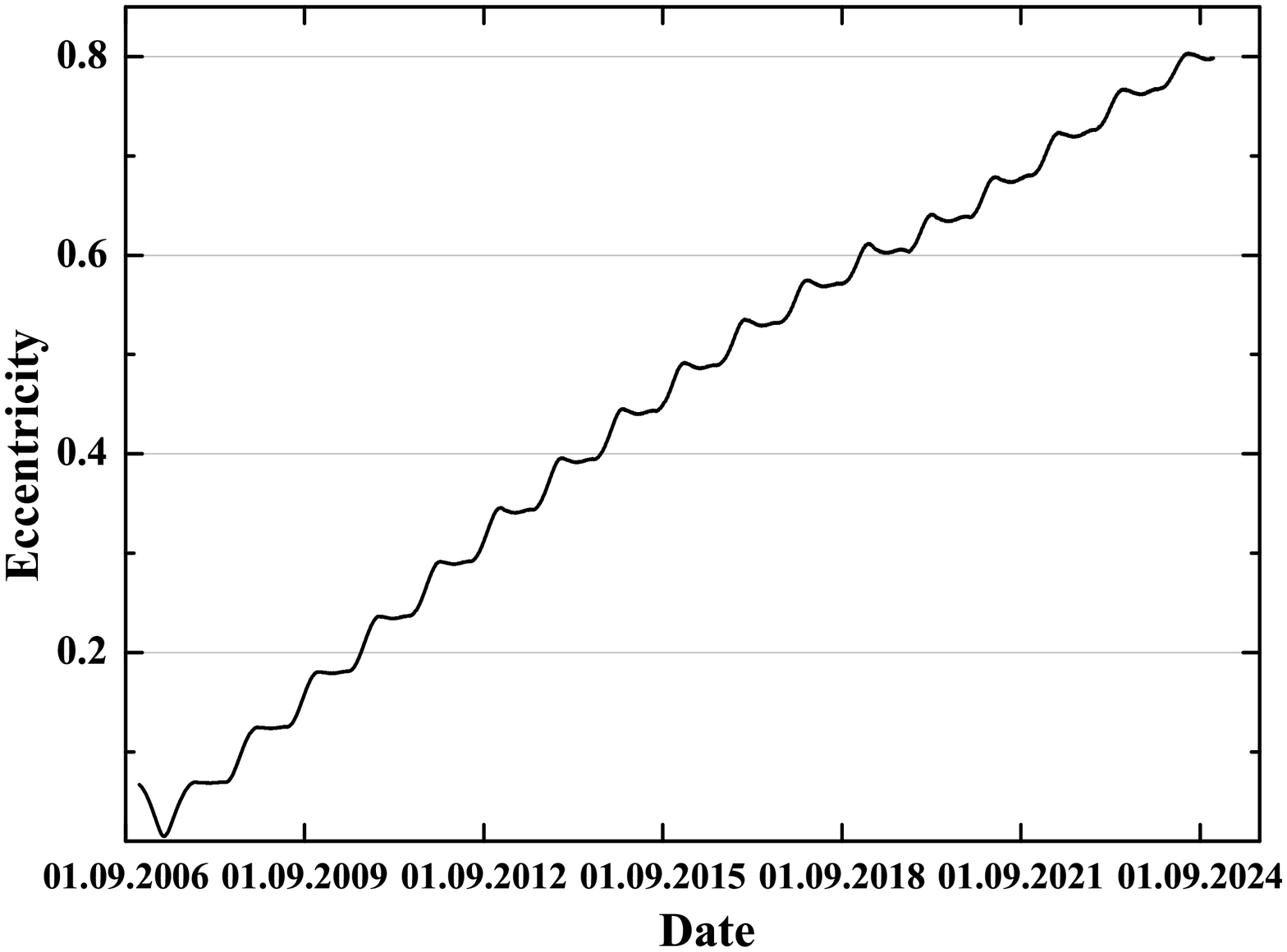}}
\resizebox{70 mm}{45 mm}
{\includegraphics {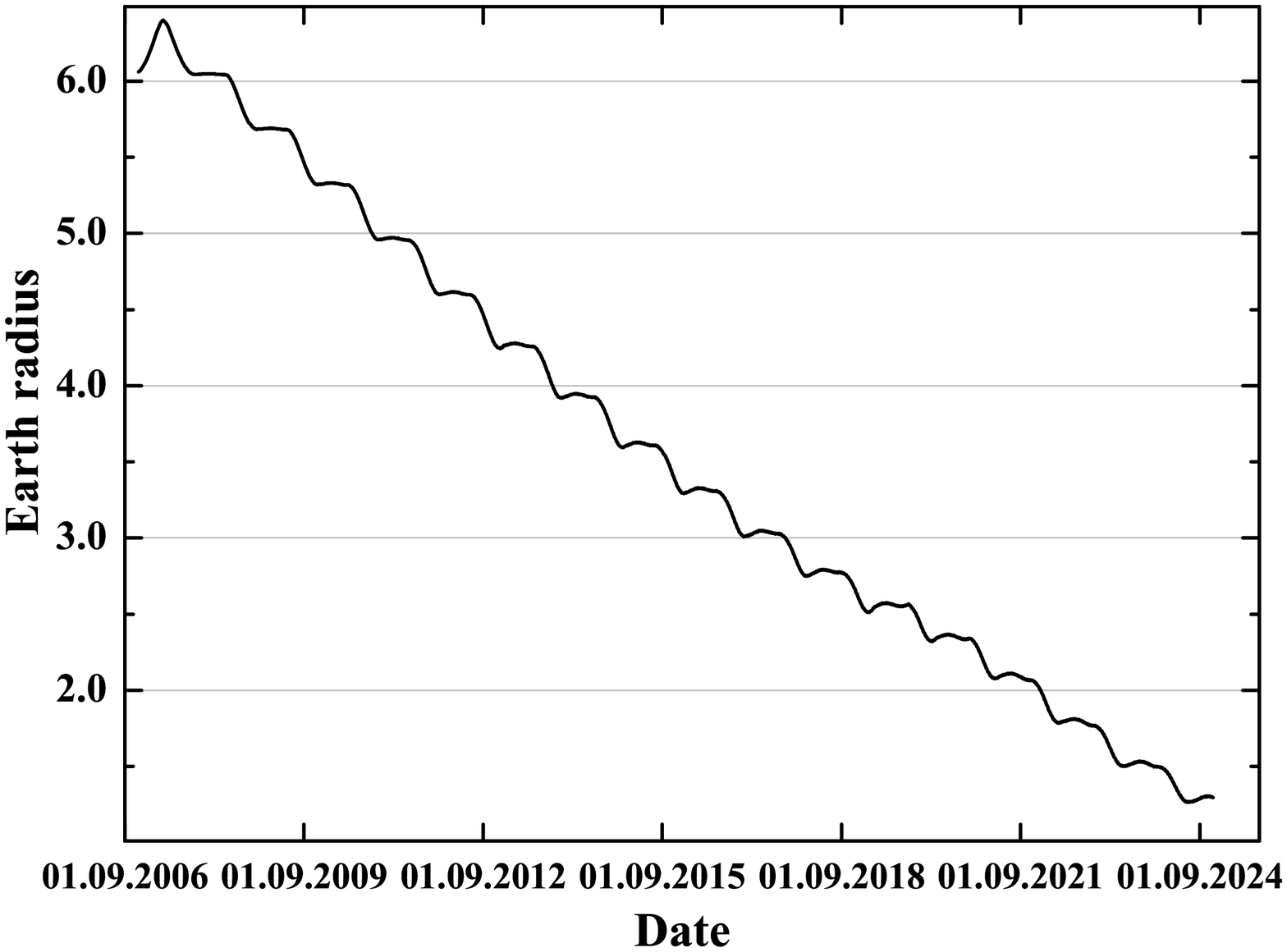}}
\caption{The change in the orbital elements of the celestial body subjected to the alternating radiation pressure.}
\label{fig3}
\end{figure}

The state vector was determined as of 24 November 2006 01:55:50.76  UTC:
\\
x = --2.28181 equatorial radius
\\
y = 6.21066  equatorial radius
\\
z = 0.54755  equatorial radius 
\\
Vx = --0.0259860  equatorial radius per minute
\\
Vy= --0.0111238 equatorial radius per minute 
\\
Vz = --0.00394734  equatorial radius per minute
\\
The indicated values were assumed to be the initial conditions for the fragment's orbit integration. 
\\
The area-to-mass ratio was assumed equal to $\frac{S}{m}=2.56$ sq.m/kg (Fr\"{u}h \& Schildknecht, 2012). The acceleration 
due to the direct solar radiation was estimated as follows:
\\

\begin{equation}
\vec{a}=C\frac{S}{m}(\frac{r_{0}}{r})^2\frac{\vec{r}-\vec{r_{S}}}{r}
\end{equation}

with $C=P_{0}(1+A)$, $P_{0} = 0.0000045606$ N/sq.m – the solar radiation pressure at the Earth's orbit, 
$A$ – the electromagnetic radiation reflection coefficient ($0<A<1$), $r_{0}$ – the average radius of 
the Earth's orbit, $\vec{r}$, $\vec{r_{S}}$  - the fragment's and the Sun's positions in the Earth-centred coordinate system.
\\
As fragment 43096 is referred as a high-orbit artificial Earth's satellites, the perturbations of its motion 
due to the Moon and Sun are comparable to those by the Earth's flattening (Borodovitsyna \& Avdyushev, 2007). In turn, the perturbations 
due to the Earth's flattening are considerably more powerful than any perturbations by all the other geopotential 
asymmetries (Borodovitsyna \& Avdyushev, 2007). Therefore, when integrating the motion equations, we accounted for 
the perturbations by the second zonal harmonic of the Earth's gravitational field, the Moon and Sun, as well as the 
solar radiation pressure. The Moon's and Sun's positions were adopted from the numerical theory DE405 
\footnote{ssd.jpl.nasa.gov}. The integration was performed by the Runge-Kutta methods of the 10th order (Bazyey \& Kara, 2005) 
during the period from 24 November 2006 to 31 July 2012. The results are presented in Figures \ref{fig1} a-d. The orbital 
element observation values are marked as dots, the solid line is resulted from the integration.
\\
\\
During the whole period the orbit's semi-major axis has not been subjected to the secular perturbations. The 
eccentricity and argument of perigee are exposed to the periodic perturbations with duration of some 370 days. 
The eccentricity varies from 0.017 to 0.071. The apse line oscillates with amplitude of about 80$^\circ$ and 
slowly rotates with the angular velocity of 0.020$^\circ$/day. The longitude of the ascending node and inclination 
of the orbit decrease at the rate of 0.0028$^\circ$/day and 0.0016$^\circ$/day, respectively, within the whole 
observation interval.
\\
Therefore, the 43096 fragment orbit periodically changes the shape and position of the apse line, leaving its 
size unaltered. Besides, the apse line, longitude of the ascending node and inclination change monotonically.
\\
The numerical integration shows that the periodic perturbations in the eccentricity and argument of perigee are 
due to the solar radiation pressure: assuming $P_{0}=0$, those perturbations disappear. Those orbital elements 
computed by the observations with the least absolute errors of $\Delta~e<0.002$ are presented in Figure \ref{fig2}. 
The dashed line corresponds to the values computed with allowance for all above-indicated perturbations. 
The solid line represents the same orbital elements computed not accounting for the radiation pressure.
\\
\\
That fact can be used to purposely change orbits of the geostationary objects and their de-orbiting to lower 
altitudes as down as the Earth's atmosphere.
\\
Let us explain that by exemplifying simulation of the 43096 fragment motion. Using the eccentricity variation curve, 
it is easy to detect that the eccentricity was increasing from 05 May 2007 to 13 November 2007, from 08 May 2008 to 
19 November 2008, from 15 May 2009 to 26 November 2009, from 22 May 2010 to 03 December 2010, and from 29 May 2011 
to 04 December 2011. During those periods the perigee distance decreases due to the radiation pressure with the 
semi-major axis remaining altered. The eccentricity was decreasing from 13 November 2007 to 08 May 2008, from 19 
November 2008 to 15 May 2009, from 26 November 2009 to 22 May 2010 and from 03 December 2010 to 29 May 2011. If 
the solar radiation pressure force is stronger, while the eccentricity increases comparing to those time intervals 
when it decreases, then it is possible to determine general secular increase of the orbital eccentricity. The same 
effect can be reached, for instance, by increasing the area-to mass ratio of the fragment while the eccentricity 
increases.
\\
We conducted the numerical experiment on the simulation of the 43096 fragment orbit with the same initial conditions 
as of 24 November 2006 01:55:50.76 UTC, but with alternating radiation pressure. It was assumed that $P_{0} = 
0.0000045606$ N/sq.m for the increasing eccentricity and $P_{0}=0$ for the decreasing eccentricity. The result is 
shown in Figure \ref{fig3}. The eccentricity increased from 0.02 as of 08 May 2007 to 0.30 as of 31 July 2012. And the 
semi-major axis remained unaltered at that. At the end of the integration interval the perigee distance decreased 
down to 29000 km (the Earth's equatorial radius 4.54). Such a considerable change in the fragment's orbit was 
successfully attained just by changing the solar radiation pressure force two times per year.
\\
\\
Colligating the result obtained, it should be noted that such method of changing the celestial body orbit in the 
near-Earth space can be applied to solve problems of the near-space ecology. Provided the capabilities to control 
the changes in the area-to-mass ratios of worn-out satellites, it is possible to solve the problem of cleaning up 
the near-Earth space from the space debris of artificial origin using the solar radiation pressure exclusively.

Thus, we processed the positional observation data for the artificial celestial body 43096, which had been obtained 
during 2006-2012 by the network of observation stations participated the ISON project. The Keplerian elements were 
determined for each series of observations, and the errors in those determinations were estimated. The state vector 
for fragment 43096 was determined as of 24 November 2006 01:55:50.76 UTC, and that value served as initial condition 
for the integration of differential motion equations in Cartesian coordinate system. The numerical integration was 
performed accounting for the perturbations due to the polar flattening of the Earth, Moon and Sun, as well as the 
solar radiation pressure. The received numerical solution is in good agreement with the observation data and 
represents periodic and secular changes of the orbit in the near-Earth space.
\\
Based on the numerical model of motion in the near-Earth space that accounts for only the most powerful 
perturbations, a new method for de-orbiting artificial celestial bodies from high altitudes is suggested.
\\
\section{Conclusion}
For the first time such a considerable amount of data over long time intervals was gathered for the objects 
with high area-to-mass ratios that enabled us to determine and estimate their observation and orbital 
characteristics. Altogether the Keldysh Institute of Applied Mathematics of the Russian Academy of Sciences 
database contains data for 247 geostationary and geostationary transiting objects with high area-to-mass ratios, 
as well as for 23 objects in the high-elliptic orbits (only objects with the results of observation for more than 
2 nights were taken into account). The number of detected relatively bright objects (which are brighter than 
magnitude 15.5) has been continuously increasing, and that is a rather surprising fact with regard to the 
continuous series of geostationary observations, which have been conducted by the ISON network for several 
years already. It is about 5-10 new objects discovered every month. Many of those new objects cross the GEO 
protected zone thereby increasing the predicted hazard for the working satellites. It is very important to 
detect as many space debris as possible to determine the sources of its origin. It is anticipated that there 
are at least several hundreds of space debris fragments brighter than magnitude 18 in the geostationary ring. 
Meanwhile, the number of weaker (and correspondingly smaller) objects can not be correctly estimated.
\\
Described here method of the celestial body orbit changing in the near-Earth space can be useful in solution 
of the near-space ecology problem, namely in the cleaning up the near-Earth space from the artificial space 
debris using the solar radiation pressure only.
\\
\acknowledgements
The authors express their gratitude to S.M. Andrievsky, the Director of the 
Astronomical Observatory of I.I. Mechnikov Odessa National University, for his valuable assistance 
in performing this study.
\\

\end{document}